\documentstyle[12pt]{article}
\begin{document}

\centerline{\LARGE Feynman Identity: \vspace{10mm} a special case. II.}
\centerline{\large G.A.T.F.da 
Costa$^a$  
}
\centerline{\large Departamento de Matem\'{a}tica}
\centerline{\large Universidade Federal de Santa Catarina}
\centerline{\large 88040-900-Florian\'{o}polis-SC-\vspace{5mm}Brasil}
\centerline{\large \vspace{5mm} and}
\centerline{\large J.Variane Jr.}
\centerline{\large Instituto de F\'{\i}sica Gleb Wataghin}
\centerline{\large Universidade Estadual de Campinas}
\centerline{\large 13083-970-Campinas-SP-Brasil}

\begin{abstract}

In this paper, the results of part I regarding a special case of
Feynman identity are extended.
The sign rule for a path in terms of data encoded by
its word and formulas for
the numbers of distinct equivalence classes of nonperiodic paths of given
length with positive or negative sign are obtained for  this extended
case. Also, a connection is found between these numbers and the generalized
Witt formula for
the dimension of certain  graded Lie algebras.
Convergence of the infinite product in the identity is proved.

\vspace{9cm}
\end{abstract}

$\overline{ e - mail \hspace{1mm} adress: gatcosta@mtm.ufsc.br}$

\newpage
\section*{I. Introduction}
 
This paper is the sequel to  Ref. 1 where 
a special case
of Feynman identity is investigated. This identity
relates admissible graphs and classes of nonperiodic paths on a lattice
and it is relevant in a combinatorial proof of Onsager's
closed formula for the partition function of the two dimensional free field
Ising model$^{1-6}$. 
In the special case considered in$^{1}$ and  here
the lattice $G_{R}$ consists of $R$ oriented
 loops and one site as shown in Figure 1.
The case $R=1$ is trivial because only one non periodic path
is possible. The case $R=2$ is the first nontrivial one and was
investigated in Ref. 1.

\begin{picture}(110,150)(-120,-30)
\put(50,50){\circle*{5}}
\put(50,17){\oval(10,40)[b]}
\put(83,50){\oval(40,10)[r]}
\put(50,83){\oval(10,40)[t]}
\put(17,50){\oval(40,10)[l]}
\put(45,17){\line(1,6){5}}
\put(55,17){\line(-1,6){5}}
\put(55,12){\vector(0,1){5}}
\put(83,55){\line(-6,-1){30}}
\put(88,55){\vector(-1,0){5}}
\put(83,45){\line(-6,1){30}}
\put(45,83){\line(1,-6){5}}
\put(45,88){\vector(0,-1){5}}
\put(55,83){\line(-1,-6){5}}
\put(17,55){\line(6,-1){30}}
\put(17,45){\line(6,1){30}}
\put(12,45){\vector(1,0){5}}
\put(60,0){1}
\put(88,60){2}
\put(63,95){3}
\multiput(30,70)(3,3){3}{\circle*{2}}
\put(2,30){$R$} 
\put(-10,-20){FIG. 1. The lattice $G_{R}$ .}
\end{picture}

In the present paper, the case with $R > 2$ will be considered. 
In this case Feynman identity
can be expressed as:
$$
\prod_{m_{1},...,m_{R} \geq 0}
(1 + z_{1}^{m_{1}}...z_{R}^{m_{R}})^{N_{+}(m_{1},...,m_{R})}
(1 - z_{1}^{m_{1}}...z_{R}^{m_{R}})^{N_{-}(m_{1},...,m_{R})}
=
\prod_{1 \leq j \leq R}(1 + z_{j})
$$
$$
\eqno(1.1)
$$
The exponents $N_{+}(m_{1},...,m_{R})$ and $N_{-}(m_{1},...,m_{R})$
are the numbers of distinct nonperiodic paths
with positive and negative signs, respectively, which
traverse $m_{1}$ times loop $1$, $m_{2}$ times loop $2$, ..., $m_{R}$
times loop $R$.
Sequences
with $m_{i} > 1$ and $m_{j \neq i}=0$ are exluded in (1.1)
because they  correspond to periodic paths.

On the left hand side of (1.1)
there are sequences
$$
S'=\{(m_{1},...,m_{R}) \mid m_{i}=1,m_{j \neq i}=0,i=1,...,R\}
\eqno(1.2)
$$ 
with $N_{+}=1$ and
$N_{-}=0$.
Collecting these sequences  we get exactly the right hand side of (1.1)
and  relation (1.1) can be equivalently expressed as
$$
\prod_{S \neq S'}
(1 + z_{1}^{m_{1}}...z_{R}^{m_{R}})^{N_{+}}
(1 - z_{1}^{m_{1}}...z_{R}^{m_{R}})^{N_{-}}
=1
\eqno(1.3)
$$
Write the product on the left hand side of (1.3)
in the following way, namely,
$$
\prod_{r=2}^{R}\prod_{G_{r}}\prod_{S(G_{r})}
(1 + z_{i_{1}}^{m_{i_{1}}}...z_{i_{r}}^{m_{i_{r}}})^{\theta_{+}(m_{i_{1}},...,
m_{i_{r}})}
(1 - z_{i_{1}}^{m_{i_{1}}}...z_{i_{r}}^{m_{i_{r}}})^{\theta_{-}(m_{i_{1}},...,
m_{i_{r}})}
\eqno(1.4)
$$
The second product runs over all subdiagrams $G_{r}$
 of ${G}_{R}$
 with $r$ loops $i_{1}, i_{2}, ..., i_{r}$, $2 \leq r \leq R$.
The third product is over all sequences 
$$
S(G_{r})=\{(m_{i_{1}},
...,m_{i_{r}}),m_{i}>0\}
\eqno(1.5)
$$
and $\theta_{+}(m_{i_{1}},...,m_{i_{r}})$
and $\theta_{-}(m_{i_{1}},...,m_{i_{r}})$
are the numbers of distinct classes of equivalence of nonperiodic paths 
with positive and negative
signs, respectively, which traverse $G_{r}$.

Given any $G_{r}$, the paths over $G_{r}$
can be classified according to the  number 
$N =m_{i_{1}}+...+m_{i_{r}}$ where $m_{i}$
is the number of times loop $i$
is traversed by a path. Thus,
$$
\prod_{S(G_{r})}
(1 + z_{i_{1}}^{m_{i_{1}}}...z_{i_{r}}^{m_{i_{r}}})^{\theta_{+}}
(1 - z_{i_{1}}^{m_{i_{1}}}...z_{i_{r}}^{m_{i_{r}}})^{\theta_{-}}
$$
$$
=
\prod_{N=r}^{\infty} 
\prod_{\begin{array}{c}
      \footnotesize{m_{i} >0}\\
       \footnotesize{m_{i_{1}}+...+m_{i_{r}}=N}
       \end{array}}
(1 + z_{i_{1}}^{m_{i_{1}}}...z_{i_{r}}^{m_{i_{r}}})^{\theta_{+}}
(1 - z_{i_{1}}^{m_{i_{1}}}...z_{i_{r}}^{m_{i_{r}}})^{\theta_{-}}
\eqno(1.6)
$$
It shall be proved that the infinite product given by relation (1.6)
converges to $1$ if $|z_{i}|<1$. This is true
for each $G_{r}$.
The proof is carried
out in Section 3  
using certain relations satisfied
by the numbers $\theta_{\pm}$, to be computed in the same section.
In Section 4, using results from section 3, a statement is proved according to
which the number of classes of nonperiodic paths of length $N$ which
traverse $G_{R}$  counterclockwisely is given by Witt formula.
Witt formula gives the dimension of certain vector spaces associated
with finite Lie algebras. Then, based on recent results in this field
 a connection is found 
%between the numbers $\theta_{+}$ and 
with the so called
generalized Witt formula for the dimension of 
certain finite graded Lie algebras.
In Section
2, a general formula for the sign of a path is obtained which is
important in Section 3 to find the number of classes of nonperiodic paths. 

\newpage
\section*{II. Rule of signs}

In this section, a general formula for the sign of any path $p$ over
$G_{R}$ is computed. It is crucial in the calculation of the numbers
$\theta_{\pm}$.

Following the methods of Ref. 1   a
path $p$ over $G_{R}$ is given by a word. This is to be understood
as an ordered
sequence of letters $D_{i}^{e_{i}}$ where $i$ gives 
the loop  of ${G}_{R}$  traversed by 
$p$ and $\mid e_{i} \mid =m_{i}$, how many times.
The sign of $e_{i}$ 
indicates 
whether the loop is traversed following the 
direction assigned for it (in this case, the
sign is positive ) or the opposite direction (in this case, the
sign is negative). A typical word is the following:
$$
{\cal W}(p)= D_{i_{1}}^{e_{i_{1}}}
             D_{i_{2}}^{e_{i_{2}}}...D_{i_{l}}^{e_{i_{l}}}
\eqno(2.1)
$$
where $l=r,r+1,...,N$. The order in which the letters
appear in the word is important since it indicates the loops  traversed
by $p$ and in which order. The order is
encoded in the sequence $S_{l}=(i_{1},i_{2},...,i_{l})$. The sequence
$S_{l}$ is such that a loop $i$ appears at least once in the sequence,
$i_{k}\neq i_{k+1}$ and $i_{l}\neq i_{1}$.
The word representation seems to imply that line $i_{1}$
is special over the others as the one to be traversed first.
This is  a convention
because the path is closed. After traversing $i_{l}$ the path joins
$i_{1}$. Fix $i_{1}=1$, from now on.

Given the word (2.1) the sequence  $(i_{1},i_{2},...,i_{l})$
can be decomposed into subsequences defined as follows. A subsequence
is an ordered set of numbers formed in such a way that
{\bf a)} if
$i,j$ are two elements inside the subsequence
and $j$ comes after $i$
then $j > i$;
${\bf b)}$ a new subsequence begins whenever this ordering is broken
 in the sequence by two adjacent elements
not satisfying {\bf a)}. For instance, a sequence with $l=14$
for the case $r=3$ is decomposed as follows:
$$
(12123213232312)=(12)(123)(2)(13)(23)(23)(12)
\eqno(2.2)
$$
Denote by $T$ the number of subsequences in a decomposition
and
let $t$ be the number of subsequences where the number $1$ is not
present
. In the example above $T=7$ and there are three subsequences 
where the number $1$ is not present, namely,
the subsequence (2) and  (23) which appears twice. So, $t=3$.

Let $n_{i}$ 
be the number of times 
line $i$ occur in the sequence $(i_{1},...,i_{l})$ and $N_{i}$,
the number of times line
$i$
is covered by $p$. Define
$$
n=n_{2}+n_{3}+...+n_{r}
\eqno(2.3)
$$
and
$$
N=N_{1}+N_{2}+...+N_{r}
\eqno(2.4)
$$
In (2.3), the sum really begins with $n_{2}$. For instance, in the word
$ D_{1}^{3} D_{2}^{2}D_{1}^{1}D_{3}^{2}D_{2}^{3}$, $n_{2}=2$, $n_{3}=1$,
$N_{1}=4$, $N_{2}=5$ and $N_{3}=2$.

The following result can now be stated:
\\
\\
\\
{\bf Lemma 1:} Given the word
 ${\cal W}(p)$ for $p$, the sign of $p$ is given by:
$$
(-1)^{N+n+s+t+1}
\eqno(2.5)
$$
\noindent {\bf Proof:}
Following Ref. 1,
relation (2.5) can be derived from the representation
of $p$ in terms of an appropriate closed normal plane curve
compatible with ${\cal W}(p)$   which winds the
loops of ${G}_{R}$ according to some rules. The
sign of $p$ comes from the number $V$ of 
selfcrossings of
its normal curve. The rules fix the way we draw a curve. Then, 
just by looking at it we  count the number $V$.

In Ref. 1, the case of a lattice with only two loops was
considered. There the selfcrossings of a normal
closed curve for $p$
were of two types:

\noindent{\bf 1) Type-1 crossings.} Crossings produced by the 
segments of the  curve winding the loops of ${G}_{R}$ like
the ones shown in Figs. 2 and 3;

\noindent{\bf 2) Type-2 crossings. } Crossings produced
when the curve changes in direction in order to wind a loop
in the opposite direction to that fixed for it like the ones
shown in Fig. 4.

In the case of a lattice with more than 
two loops, a third type of crossings is possible to occur:

\noindent{\bf 3) Type-3 crossings. } Crossings
produced when segments linking different loops of
${G}_{R}$
intersect.

\noindent{\bf Counting Type-1 crossings.}
According to the rules given in Ref. 1
the segment of curve
 $D_{i}^{x}$
is drawn making an inward spiral around loop
 $i$. The number of times loop $i$ is traversed
by this segment is $\mid x \mid$.
If $x>0$, then the curve winds the
 loop following its orientation. If $x<0$, then it winds 
$i$ with opposite 
orientation.
At the end of the winding when the curve leaves loop $i$, it will cross
itself a number of times given by
$\mid x \mid -1$.

Let's first consider the case where all exponents are positive,
$e_{i} >0$.
Denote by $e_{1,\alpha}$, $\alpha=1,2,...,n_{1}$, the exponents
of $D_{1}$ in  ${\cal W}(p)$ where $n_{1}$ is the number
of times loop $i=1$ appears in the sequence $(i_{1},...,i_{l})$.
Fig. 2  shows the sequence of segments
$$
D_{1}^{e_{11}},D_{1}^{e_{12}},...,D_{1}^{e_{1n_{1}}}
\eqno(2.6)
$$
The number of crossings in this case is (see Ref. 1, Section 2, for more
details)
$$
A_{1}=(e_{11}-1)+2(e_{12}-1)+...+2(n_{1}-1)(e_{1n_{1}}-1)
\eqno(2.7)
$$

\begin{picture}(200,300)(-70,-215)

\put (61,34){\vector(0,-1){10}}
\put (61,-1){\line(0,1){50}}
\put (136,20){\vector(0,1){10}}
\put (136,-7){\line(0,1){55}}

\put (-3,-1){\line(3,0){64}}
\put (-1,-3) {\line(3,0){202}}
\put (1.5,-5) {\line(3,0){197}}
\put (136.5,-7){\line(3,0){60.5}}

\put (-3,-1){\line(0,-1){175}}
\put (-1,-174) {\line(0,1){171}}
\put (1,-172) {\line(0,1){167}}

\put (-3,-176) {\line(3,0){203.5}}
\put (-1,-174) {\line(3,0){200}}
\put (1,-172) {\line(3,0){196}}

\put (201,-176) {\line(0,1){173}}
\put (199,-174) {\line(0,1){169}}
\put (197,-172) {\line(0,1){165}}

\put (210,-100){$D_{1}^{e_{1n_{1}}}$}
%***
\put (77,34){\vector(0,-1){10}}
\put (77,-23){\line(0,1){72}}
\put (120,20){\vector(0,1){10}}
\put (120,-29){\line(0,1){78}}

\put (24,-23){\line(3,0){53}}
\put (26,-25){\line(3,0){148}}
\put (28,-27){\line(3,0){144}}
\put (120,-29){\line(3,0){50}}

\put (24,-23){\line(0,-1){131}}
\put (26,-25){\line(0,-1){127}}
\put (28,-27){\line(0,-1){122.5}}

\put (27.5,-150) {\line(3,0){143}}
\put (25.5,-152) {\line(3,0){147}}
\put (23.5,-154) {\line(3,0){151.5}}

\put (175,-154) {\line(0,1){129}}
\put (173,-152) {\line(0,1){125}}
\put (171,-150.5) {\line(0,1){121}}

\put (176,-100){$D_{1}^{e_{13}}$}
\put (180,-110){...}

%****

\put (85,-49){\line(0,1){98}}
\put (85,34){\vector(0,-1){10}}
\put (111,-55.5){\line(0,1){104}}
\put (111,20){\vector(0,1){10}}

\put (50,-49){\line(3,0){34.5}}
\put (52.5,-51){\line(3,0){93.5}}
\put (54.5,-53){\line(3,0){89}}
\put (111,-55){\line(3,0){30.5}}

\put (50,-132.5) {\line(0,1){82.5}}
\put (52,-130) {\line(0,1){79}}
\put (54,-128) {\line(0,1){75.5}}

\put (54,-128){\line(3,0){88}}
\put (52,-130){\line(3,0){92.5}}
\put (50,-132){\line(3,0){96.5}}

\put (146,-132) {\line(0,1){81}}
\put (144,-130) {\line(0,1){77}}
\put (142,-128) {\line(0,1){73.5}}

\put (148,-100){$D_{1}^{e_{12}}$}

%***

\put (95,-75){\line(0,1){10}}
\put (95,-66){\circle*{3}}
\put (95,-70){\vector(0,-1){5}}
\put (102,-81.5){\line(0,1){130}}
\put (102,20){\vector(0,1){10}}

\put (80,-75){\line(3,0){15}}
\put (82,-77){\line(3,0){41}}
\put (84,-79){\line(3,0){36.5}}
\put (102.5,-81){\line(3,0){15.5}}

\put (80,-110) {\line(0,1){35}}
\put (82,-108) {\line(0,1){31}}
\put (84,-106) {\line(0,1){27}}

\put (80,-110){\line(3,0){43}}
\put (82,-108){\line(3,0){38.5}}
\put (84,-106){\line(3,0){35}}

%%%
\put (119,-106.5) {\line(0,1){26}}
\put (121,-108) {\line(0,1){29}}
\put (123,-110) {\line(0,1){33}}

\put (124,-100){$D_{1}^{e_{11}}$}
\put (85,-70){e}

\put (0,-210){FIG. 2. How to count type-1 crossings .(I).}

\end{picture}

The other loops $i,i=2,...,r$, (see Fig. 3), contribute with
$$
B_{i}=(e_{i1}-1)+3(e_{i2}-1)+5(e_{i3}-1)+...+(2n_{i}-1)(e_{in_{i}}-1)
\eqno(2.8)
$$
After winding loop $i_{l}$, the curve 
goes to its point of ``departure''
 at $e$ (see Fig. 2) but to do so it has to 
cross all the segments which have already winded loop $i_{1}=1$.
The number of these crossings is given by
$$
C=e_{12}+...+e_{1n_{1}}
\eqno(2.9)
$$
Therefore, the total number of crossings  is
$$
V=A_{1} + \sum_{j=2}^{r} B_{j} + C
\eqno(2.10)
$$
A simple calculation shows that in this case
$$
(-1)^{V}=(-1)^{N+n+1}
\eqno(2.11)
$$

\begin{picture}(200,300)(-81,-270)

\put (67,-220){\line(0,1){50.5}}
\put (67,-186){\vector(0,-1){10}}
\put (136,-220){\line(0,1){43}}
\put (136,-200){\vector(0,1){10}}

\put (-1,-3) {\line(3,0){202}}
\put (1,-5) {\line(3,0){197.5}}
\put (3,-7){\line(3,0){194}}

\put (-1,-174) {\line(0,1){171}}
\put (1,-172) {\line(0,1){167}}
\put (3,-8){\line(0,-1){162.5}}

\put (3,-170) {\line(3,0){64.5}}
\put (1,-172) {\line(3,0){196.5}}
\put (-1,-174) {\line(3,0){200}}
\put (136,-176) {\line(3,0){64.5}}

\put (201,-176.5) {\line(0,1){173.5}}
\put (199,-174.5) {\line(0,1){169.5}}
\put (197,-172) {\line(0,1){165}}

\put (210,-100){$D_{i}^{e_{i_{k_{i}}}}$}
%***

\put (79.5,-220){\line(0,1){72}}
\put (79.5,-186){\vector(0,-1){10}}
\put (125.5,-220){\line(0,1){66}}
\put (125.5,-200){\vector(0,1){10}}

%\put (24,-23){\line(3,0){53}}
\put (26,-25){\line(3,0){148}}
\put (28,-27){\line(3,0){144}}
\put (30,-29){\line(3,0){140}}

\put (26,-25){\line(0,-1){127}}
\put (28,-27){\line(0,-1){122.5}}
\put (30,-29){\line(0,-1){120}}

\put (29.5,-148) {\line(3,0){50}}
\put (27.5,-150) {\line(3,0){143}}
\put (25.5,-152) {\line(3,0){147}}
\put (126,-154) {\line(3,0){50}}

\put (175,-154) {\line(0,1){129}}
\put (173,-152) {\line(0,1){125}}
\put (171,-150.5) {\line(0,1){121}}

\put (176,-100){$D_{i}^{e_{i_{3}}}$}
\put (180,-110){...}

%****

\put (92,-220){\line(0,1){93}}
\put (92,-186){\vector(0,-1){10}}
\put (114,-220){\line(0,1){88}}
\put (114,-200){\vector(0,1){10}}

%%\put (50,-49){\line(3,0){34.5}}
\put (52,-51){\line(3,0){94}}
\put (54,-53){\line(3,0){89.5}}
\put (56,-55){\line(3,0){85}}

\put (52,-130.5) {\line(0,1){79}}
\put (54,-128.5) {\line(0,1){75.5}}
\put (56,-126) {\line(0,1){70}}

\put (56,-126){\line(3,0){35.5}}
\put (54,-128){\line(3,0){88}}
\put (52,-130){\line(3,0){92}}
\put (115,-132){\line(3,0){30}}
%*******

\put (146,-132.5) {\line(0,1){81}}
\put (144,-130) {\line(0,1){76.5}}
\put (142,-128) {\line(0,1){72.5}}

\put (148,-100){$D_{i}^{e_{i_{2}}}$}

%***

\put (99,-220){\line(0,1){116}}
\put (99,-186){\vector(0,-1){10}}
\put (107,-220){\line(0,1){110}}
\put (107,-200){\vector(0,1){10}}

%\put (80,-75){\line(3,0){15}}
\put (82,-77){\line(3,0){41}}
\put (84,-79){\line(3,0){36.5}}
\put (86,-81){\line(3,0){33}}

\put (82,-108) {\line(0,1){31}}
\put (84,-106) {\line(0,1){27}}
\put (86,-104) {\line(0,1){23}}

\put (86,-104){\line(3,0){12}}
\put (82,-108){\line(3,0){38.5}}
\put (84,-106){\line(3,0){35}}
\put (107,-110){\line(3,0){15.5}}

\put (119,-106.5) {\line(0,1){25}}
\put (121,-108) {\line(0,1){29}}
\put (123,-110) {\line(0,1){33}}

\put (124,-100){$D_{i}^{e_{i_{1}}}$}

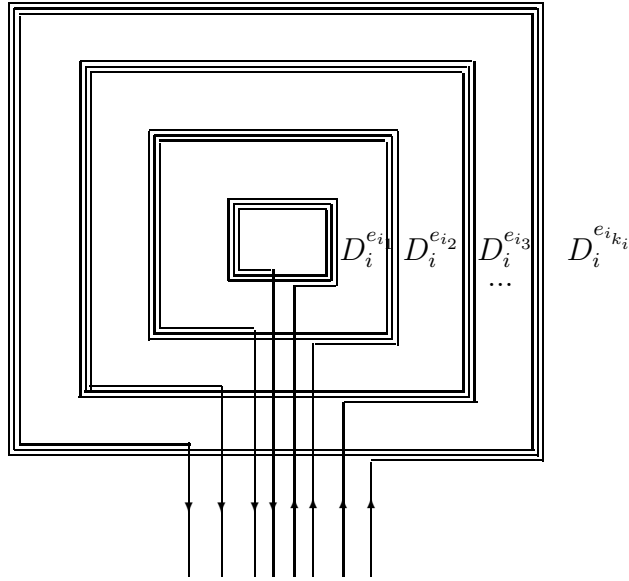
\put (0,-260){FIG. 3. How to count type-1 crossings.(II).}

\end{picture}

\noindent {\bf Counting type-2 crossings.}
In order to find (2.11) 
we considered words with all the exponents $e_{i}$
positive. In general, however, they may be negative, too. So, now, we
should consider this
more general case.

Before winding a loop the curve approaches it
 from 
down the right. See Fig. 4. If $e_{i}>0$ it goes on to wind the loop
 counterclockwisely
$\mid e_{i} \mid$ times leaving in the end by the left side.
When $ e_{i} < 0$, the curve coming from down the right
will first turn to the left and only  then will go upward to
wind the loop $\mid e_{i} \mid$ times
clockwisely crossing itself once on its way
out. The second time the same loop is winded counterclockwisely, the
curve will cross itself five times on its way out. See Fig. 5.

\begin{picture}(70,150)(-90,-30)
\put (50,15){\line(0,1){85}}
\put (140,8){\line(0,1){92}}
\put (50,14){\line(8,0){72}}
\put (122,-6){\line(0,1){20}}
\put (50,100){\line(25,0){91}}
\put (50,8){\line(25,0){91}}
\put (50,-8){\line(0,1){15}}
\put(50,0){\vector(0,-1){5}}
\put(122,0){\vector(0,1){5}}
\put(95,35){\circle*{5}}
\put(95,68){\oval(10,40)[t]}
\put (93,70){$i$}
\put(90,68){\line(1,-6){5}}
\put(90,73){\vector(0,-1){5}}
\put(100,68){\line(-1,-6){5}}
\put(-20,-40){FIG. 4. How to count type-2 crossings . (I) .}
\end{picture}

If there are $s_{i}$ occurrences of loop $i$ in the
sequence $(i_{1},...,i_{l})$ for ${\cal W}$ 
with $e_{i} < 0$ the curve will cross itself
$$
V_{i}=\sum_{x=1}^{s_{i}}(4x-3)
\eqno(2.12)
$$
times.

\begin{picture}(200,250)(-70,-150)

\put (50,-35){\line(0,1){85}}
\put (140,-42){\line(0,1){92}}
\put (50,-35){\line(8,0){72}}
\put (122,-95){\line(0,1){60}}
\put (50,50){\line(25,0){91}}
\put (50,-42){\line(25,0){91}}
\put (50,-95){\line(0,1){53}}
\put(50,-85){\vector(0,-1){5}}
\put(122,-90){\vector(0,1){5}}
\put(95,-15){\circle*{5}}
\put(95,18){\oval(10,40)[t]}
\put(90,18){\line(1,-6){5}}
\put(90,23){\vector(0,-1){5}}
\put(100,18){\line(-1,-6){5}}
\put (94,20){$i$}

\put (160,-95){\line(0,1){40}}
\put(160,-85){\vector(0,1){5}}
\put (30,-55){\line(25,0){130}}
\put (30,-55){\line(0,1){125}}
\put (30,70){\line(8,0){150}}
\put (180,-70){\line(0,1){140}}
\put (15,-95){\line(0,1){25}}
\put(15,-85){\vector(0,-1){5}}
%
%\put (200,-150){\line(0,1){40}}
\put (15,-70){\line(25,0){165}}

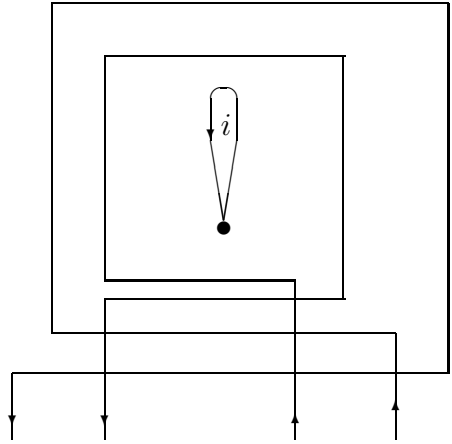
\put (-10,-130){FIG. 5. How to count type-2 crossings .(II).}

\end{picture}

The contribution of these crossings to the sign is, therefore,
$$
(-1)^{V_{i}}=(-1)^{s_{i}}
\eqno(2.13)
$$
The total contribution to the sign of $p$ coming from
$s=\sum s_{i}$
negative exponents in ${\cal W}$ is then given by
$$
(-1)^s
\eqno(2.14)
$$
and so we get
$$
(-1)^{N+n+s+1}
\eqno(2.15)
$$

\noindent{\bf Counting type-3 crossings.}
First, let's consider the case
of a sequence $(i_{1},...,i_{l})$ that has a
decomposition into $T$ subsequences all of them
 beginning with loop $1$. Moreover, suppose that the subsequences 
up to the $(T-1)$-th subsequence are of the form $(1,2,...,r)$, that is,
all loops of ${G}_{r}$ are present but there are  gaps
(the lack of one or more loops of ${G}_{r}$) inside the $T$-th subsequence.

In order to illustrate the implications
of gaps for the counting of
type-3 crossings we are going to consider the
simplest case where there is
 only one gap, that between 1 and $i_{a}$, namely,
$$
(1 \hspace{4mm} i_{a}...r)
\eqno(2.16)
$$
where $i_{a}\neq 2$ and lines $2,3,...,i_{a}-1$ are not in this
subsequence.

Having assumed that
all subsequences up to
the $(T-1)$-th have no gaps means that each of the $r$ loops of
${G}_{r}$
have been already winded before $(T-1)$ times. So, 
 when the curve goes from line 1 to line $i_{a}$ it has to cross
 twice
a bundle with  $(T-1)$ segments of itself thus
producing $2(T-1)$ crossings. See Fig. 6.

\begin{picture}(200,240)(-70,-215)

\put (290,-150){\line(0,8){75}}
\put (290, -100){\vector(0,1){5}}
\put (235,-145){\line(0,8){145}}
\put (235, -100){\vector(0,1){5}}
\put (195,-130){\line(0,8){115}}
\put (165,-130){\line(8,0){30}}
\put (150,-130){...}
\put (180, -130){\vector(1,0){5}}

\put (-50,-150){\line(8,0){340}}
\put (-50,-150){\line(0,8){140}}
\put (-50,-10){\line(8,0){25}}
\put (-25,-60){\line(0,8){50}}
\put (-25,-60){\line(8,0){6}}
\put (-13,-65){...}
\put (-45,-145){\line(8,0){280}}
\put (-45,-145){\line(0,8){130}}
\put (-45,-15){\line(8,0){14}}
\put (-30,-65){\line(0,8){50}}
\put (-30,-65){\line(8,0){12}}
\put (-38,-23){.}
\put (-38,-26){.}
\put (-38,-29){.}
\put (-13,-60){...}
\put (-40,-120){\line(8,0){210}}
\put (-40,-120){\line(0,8){79}}
\put (-40,-40){\line(8,0){4}}
\put (-35,-90){\line(0,8){50}}
\put (-35,-90){\line(8,0){15}}
\put (-13,-90){...}
\put (4,-60){\line(8,0){6}}
\put (10,-60){\line(0,8){50}}
\put (10,-10){\line(8,0){25}}
\put (35,-100){\line(0,8){90}}
\put (35,-100){\line(8,0){80}}
\put (115,-100){\line(0,8){100}}
\put (115,0){\line(8,0){120}}
\put (30,-65){\line(8,0){10}}
\put (15,-65){\line(0,8){50}}
\put (15,-15){\line(8,0){14}}
\put (30,-65){\line(0,8){50}}
\put (0,-65){\line(8,0){15}}
\put (22,-23){.}
\put (22,-26){.}
\put (22,-29){.}
\put (50,-65){...}
%**
\put (10,-90){\line(8,0){10}}
\put (20,-90){\line(0,8){50}}
\put (20,-40){\line(8,0){4}}
\put (25,-90){\line(0,8){50}}
\put (25,-90){\line(8,0){15}}
\put (50,-90){...}
%%%

%
\put (64,-65){\line(8,0){10}}
\put (75,-65){\line(0,8){50}}
\put (75,-15){\line(8,0){14}}
\put (90,-65){\line(0,8){50}}
\put (90,-65){\line(8,0){40}}
\put (65,-90){\line(8,0){15}}
\put (80,-90){\line(0,8){50}}
\put (80,-40){\line(8,0){4}}
\put (85,-90){\line(0,8){50}}
\put (82,-23){.}
\put (82,-26){.}
\put (82,-29){.}
\put (85,-90){\line(8,0){65}}
%%%%%%%%%
\put (130,-65){\line(0,8){50}}
\put (130,-15){\line(8,0){65}}
\put (170,-121){\line(0,8){86}}
\put (140,-35){\line(8,0){30}}
\put (140,-45){\line(0,8){10}}
\put (140,-38){\vector(0,-1){5}}
\put (115,-38){\vector(0,-1){5}}
\put (35,-38){\vector(0,1){5}}
\put (150,-23){.}
\put (150,-26){.}
\put (150,-29){.}
\put (140,-53){.}
\put (140,-56){.}
\put (140,-59){.}
\put (118,-60){\tiny{T-1}}
\put (120,-73){.}
\put (120,-76){.}
\put (120,-79){.}
\put (118,-95){\tiny{1}}
\put (70,-73){.}
\put (70,-76){.}
\put (70,-79){.}
\put (30,-73){.}
\put (30,-76){.}
\put (30,-79){.}

%3
\put (150,-90){\line(0,8){50}}
%2
\put (150,-40){\line(8,0){4}}
%1
\put (155,-71){\line(0,8){30}}
\put(155,-71){\circle*{2}}
\put(157,-71){e}
\put(157,-74){.........................................}
\put(155,-90){\tiny{1}}
\put(175,-90){\tiny{2}}
\put (180,-90){...}
\put(200,-90){\tiny{T-1}}
\put(245,-90){\tiny{T}}

\put(155, -50){\vector(0,1){5}}

\put (150,5){ 1 }
\put (75,5){ 2 }
\put (20,5){$i_{a}$}
\put (-40,5){r}

\put (5,-200){FIG. 6. How to count type-3 crossings. (I).}

\end{picture}

(For convenience of presentation, in Figs. 6 and 7, the curve
windings of loops have not been  displayed like in  Figs. 2 and 3.)
For the same reason, if there were other gaps,
whenever
one gap is met by the curve in the same subsequence
or in another one in the decomposition,
the total number of type-3 crossings would be an even number.
The number of type-3 crossings being an
even number their contribution to the sign of $p$
is then $+1 \equiv (-1)^{t}$ with $t=0$.

An odd number of type-3 crossings is of course
possible but only when the sequence $(i_{1},...,i_{l})$ 
is such that it has subsequences in its decomposition 
which does not initiate with
loop $1$. To see that, let's  consider a
simple representative case, namely,
the decomposition
$$
(12...r)...(12...r)(a....r)
\eqno(2.17)
$$
where $a \neq 1$
and the first  $(T-1)$
subsequences before $(a....r)$
have no gaps at all.
When the curve goes over the first $(T-1)$ subsequences 
the loops of ${G}_{r}$ are winded $(T-1)$ times each.
Then, it has to cross once 
a bundle with $(T-2)$ segments and then another bundle with $(T-1)$
 segments before
winding loop  $a$. (See Fig. 7).
The different numbers of segments in each bundle is solely
due to the fact that loop 1 was not traversed by $p$.
A total of $(2T-3)$
crossings is produced and their 
contribution to the sign is $-1 \equiv {(-1)}^{t}$ where $t=1$.
In a more general case
where there are more subsequences lacking loop $1$ exactly
$t$ times one can show that the sign is $(-1)^{t}$.
In this way we get the LHS of (2.5).
$\Box$

\begin{picture}(200,250)(-70,-215)

%1
\put (290,-150){\line(0,8){75}}
\put(290, -100){\vector(0,1){5}}
%2
\put (-60,-150){\line(8,0){350}}
%3
\put (-60,-150){\line(0,8){140}}
%4
\put (-60,-10){\line(8,0){35}}
%5
\put (-25,-60){\line(0,8){50}}
%6
\put (-25,-60){\line(8,0){6}}
%7
\put (-13,-60){...}
%8
\put (4,-60){\line(8,0){6}}
%9
\put (10,-60){\line(0,8){50}}
%10
\put (10,-10){\line(8,0){25}}
%11
\put (35,-145){\line(0,8){135}}
\put(35, -130){\vector(0,1){5}}
\put(35, -80){\vector(0,1){5}}
\put(35, -30){\vector(0,1){5}}
%2
%12
\put (-55, -145){\line(8,0){91}}
\put (-20, -145){\vector(1,0){5}}
%13
\put (-56,-145){\line(0,8){132}}
\put(-56, -40){\vector(0,-1){5}}
%14
\put (-55,-15){\line(8,0){25}}
%15
\put (-30,-65){\line(0,8){50}}
%16
\put (-30,-65){\line(8,0){12}}
%17
\put (-13,-65){...}
%18
\put (0,-65){\line(8,0){15}}
%19
\put (15,-65){\line(0,8){50}}
%20
\put (15,-15){\line(8,0){14}}
%21
\put (30,-65){\line(0,8){50}}
%22
\put (30,-65){\line(8,0){10}}
%23
\put (50,-65){...}
%24
\put (64,-65){\line(8,0){10}}
%25
\put (75,-65){\line(0,8){50}}
%26
\put (75,-15){\line(8,0){14}}
%27
\put (90,-65){\line(0,8){50}}
%28
\put (90,-65){\line(8,0){40}}
\put(110, -65){\vector(-1,0){5}}
%29
\put (130,-65){\line(0,8){50}}
%30
\put (130,-15){\line(8,0){70}}
%31
\put (200,-140){\line(0,8){125}}
\put(200, -80){\vector(0,1){5}}
%32
\put (-50,-140){\line(8,0){250}}
\put(70, -140){\vector(1,0){5}}
\put(0, -140){\vector(1,0){5}}
\put (-50,-140){\line(0,8){20}}
\put (-50,-110){.}
\put (-50, -113){.}
\put (-50, -117){.}
%33
\put (-35,-90){\line(8,0){15}}
%34
\put (-13,-90){...}
\put (-38,-23){.}
\put (-38,-26){.}
\put (-38,-29){.}
%a
\put (-40,-120){\line(8,0){210}}
\put (50,-120){\vector(1,0){5}}
%b
\put (-40,-120){\line(0,8){80}}
%c
\put (-40,-40){\line(8,0){4}}
%d
\put (-35,-90){\line(0,8){50}}
%e
\put (10,-90){\line(8,0){10}}
%f
\put (20,-90){\line(0,8){50}}
%g
\put (20,-40){\line(8,0){4}}
%h
\put (25,-90){\line(0,8){50}}
%i
\put (25,-90){\line(8,0){15}}
%j
\put (50,-90){...}
%h
\put (65,-90){\line(8,0){15}}
%i
\put (80,-90){\line(0,8){50}}
%j
\put (80,-40){\line(8,0){4}}
%k
\put (85,-90){\line(0,8){50}}
%l
\put (85,-90){\line(8,0){65}}
\put (110,-90){\vector(-1,0){5}}
%m
\put (150,-90){\line(0,8){50}}
%n
\put (150,-40){\line(8,0){4}}
%oeeeeeee
\put (155,-75){\line(0,8){35}}
\put (155,-50){\vector(0,1){5}}
%p
\put (170,-120){\line(0,8){85}}
%q
\put (140,-35){\line(8,0){30}}
\put (140,-45){\line(0,8){10}}
\put (140,-48){.}
\put (140,-51){.}
\put (140,-54){.}

\put (22,-23){.}
\put (22,-26){.}
\put (22,-29){.}

\put (82,-23){.}
\put (82,-26){.}
\put (82,-29){.}

\put (150,5){ 1 }
\put (75,5){ 2 }
\put (20,5){$i_{a}$}
\put (-40,5){r}

\put(157,-74){.........................................}
\put(155,-90){\tiny{1}}
\put(175,-90){\tiny{2}}
\put (180,-90){...}
\put(200,-90){\tiny{T-1}}

\put (5,-200){FIG. 7. How to count type-3 crossings. (II).}

\end{picture}

\noindent {\bf Corollary 1.} The sign of $p$ is given as well by
$$
(-1)^{N+l+s+T+1}
\eqno(2.18)
$$
\noindent {\bf Proof:}
Add $2n_{1}$ to the exponent of $(-1)$ in (2.5).
Then,
$$
(-1)^{N+n+s+t+1} \equiv (-1)^{N+(n_{1}+n)+s+(t+n_{1})+1}
\eqno(2.19)
$$
But $n_{1}+n \equiv l$ and $t+n_{1} \equiv T$. $\Box$

\noindent {\bf Corollary 2.} The sign of a periodic word $p$ equals
the sign of its nonperiodic subword if $p$ has odd period and it
is $-1$ if the period is an even number.

\noindent {\bf Proof:} Suppose $p$ given by the word
$$
 D_{i_{1}}^{e_{i_{1}}}
             D_{i_{2}}^{e_{i_{2}}}...D_{i_{l}}^{e_{i_{l}}}
\eqno(2.20)
$$
for a given $l \leq N$, $N =\sum |e_{i}| $, $s$ negative exponents
and $T$ subsequences in the sequence $(i_{1},...,i_{l})$.
Suppose $p$ is periodic with period $g$. 
Then, $p$  is the repetition
of a non periodic subword $g$ times, that is, $p = (w)^{g}$ where
$$
w = D_{i_{1}}^{e_{i_{1}}}
             D_{i_{2}}^{e_{i_{2}}}...D_{i_{j}}^{e_{i_{j}}}
\eqno(2.21)
$$
with $j=l/g$, length $L= N/g$, $s_{0}=s/g$ negative exponents and $T_{0}
=T/g$ subsequences in $(i_{1},...,i_{j})$. The sign of $p$ is
$$
(-1)^{N+l+s+T+1}= (-1)^{g(n+j+s_{0}+T_{0})+1}
\eqno(2.22)
$$
which equals $(-1)$ if $g$ is an even number and equals the sign
of $w$ if $g$ is an odd number. $\Box$

\newpage
\section*{III. The numbers $\theta_{\pm}$}

In this section we 
compute explicit formulas for the weights $\theta_{\pm}$ 
and prove convergence of the infinite product (1.6).

Let's consider  the simpler case $z_{1}=z_{2}=...=z_{R}=z$.
In this case (1.6) becomes
$$
\prod_{N=r}^{\infty} (1 + z^{N})^{\theta_{+}(N,r)}
(1-z^{N})^{\theta_{-}(N,r)}
\eqno(3.1)
$$
where
$$
\theta_{\pm}(N,r)=\sum_{\begin{array}{c}
      \footnotesize{m_{i} >0}\\
       \footnotesize{m_{i_{1}}+...+m_{i_{r}}=N}
       \end{array}}
\theta_{\pm}(m_{i_{1}},...,m_{i_{r}})
\eqno(3.2)
$$

\noindent{\bf Theorem 3.1.} 
Given  $r$, $N \geq r$, the number of equivalence
classes of nonperiodic paths of length $N$ and positive sign which
traverses $r$ loops of $G_{R}$ is given by
$$
\theta_{+}(N,r)=\sum_{odd \hspace{1mm} g \mid N}
\frac{\mu(g)}{g} {\cal F}_{r} \left( \frac{N}{g} \right)
\eqno(3.3)
$$
where the summation is over the odd divisors of $N$ only and
$$
2y{\cal F}_{r}(y)=
\sum_{k=-1}^{r-1}
(-1)^{r+k+1}
\left( \begin{array}{c}
                   r \\ k+1
\end{array}\right)
(2k+1)^{y}
\eqno(3.4)
$$
For the number $\theta_{-}(N,r)$ of equivalence classes of nonperiodic paths
of length $N$ and negative sign which traverses $r$ loops
of $G_{R}$, the following cases hold:

\noindent{\bf I)} If $N$ is {\bf a)} odd or prime or {\bf b)} even 
but $N < 2r$, then
$$
\theta_{-}(N,r)=\theta_{+}(N,r)
\eqno(3.5)
$$
\noindent{\bf II)} If $N$ is even and $N \geq 2r$, then
$$
\theta_{-}(N,r)=\theta_{+}(N,r) - \theta_{+}(\frac{N}{2},r)
\eqno(3.6)
$$
Furthemore,
if $\mid z \mid <1$
the product (3.1) converges to $1$.

\noindent{\bf Proof:} The proof of (3.3) follows  ideas
from
Ref. 1 and uses the general formula
for the sign of a path computed in the previous section.
The proof is lengthy and for this reason left to the Appendix.
Let's prove $I)$ and $II)$ and convergence.
In Ref. 1 relations (3.5) and (3.6) were proved for the case
$r=2$ by direct, lengthy computation.
It is possible to repeat that for the present case but
simpler arguments can be used. They are as 
follows.
A path traverses all $r$ lines of $G_{r}$
so its length is $N \geq r$. It is clear that if $N < 2r$
the path can not be periodic for to be periodic a path must have
a length which is a multiple of $r$ and period $g \geq 2$
so its length should be $N \geq 2r$. Since there is no periodic path with
length $N < 2r$ then in this case 
$\theta_{+}=\theta_{-}$. If $N$ is any prime number, the only
divisors of $N$ are $1$ and $N$, hence, there can be no periodic
paths with prime length and again $\theta_{+}=\theta_{-}$. Suppose now that $N$ is an odd but nonprime number.
Let's prove that the numbers of periodic words with $+$ and $-$ signs
are equal and therefore one must have $\theta_{+}=\theta_{-}$. Call $N_{0}$
the first odd, nonprime number greater than $2r$. If there are
odd numbers in between
$2r$ and $N_{0}$, then they must be prime and for these we have already 
proved that 
$\theta_{+}=\theta_{-}$. The divisors of $N_{0}$ are all odd.
Denote them by $1,g_{0,1},...,g_{0,m},N_{0}$. Then, there are odd numbers
$n_{0,1},...,n_{0,m}$ such that $N_{0}=n_{0,1}g_{0,1}=...=n_{0,m}g_{0,m}$.
So, the number of periodic paths with period $g_{0,i}$, $i=1,2,...,m$,
 and positive (negative)
sign is given by $\theta_{+}(n_{0,i})$ ( $\theta_{-}(n_{0,i})$).
Since $n_{0,i} < N_{0}$, $n_{0,i}$ is prime and
$\theta_{+}(n_{0,i})=\theta_{-}(n_{0,i})$.
Hence, one must have $\theta_{+}(N_{0})=\theta_{-}(N_{0})$. Induction, 
now, proves that  $\theta_{+}(N)=\theta_{-}(N)$ for any  odd and nonprime
number $N$.

Consider now the case $N \geq 2r$, and $N$ is an even number. Suppose, first, 
$r$ odd and prime.  
From Corollary 2, section 2, the sign of a periodic
word with even period is $-1$ so 
in the case the period is $g=2$, 
all the periodic words have sign $-1$. 
The number of nonperiodic
words with positive sign  and length $N/2$ which make periodic 
words of length $2r$ is  $\theta_{-}(r)$, hence, subtracting the periodic words
one gets
$\theta_{-}(2r)=\theta_{+}(2r)-
\theta_{-}(r)$. Suppose now that $r$ is odd but nonprime. 
In this case the divisors
of $N=2r$ are $1,2,a,2r$ where $a$ runs the divisors of $r$, hence, $a$
is odd. 
If
the period is $a$ then the length of nonperiodic subwords is $2r/a$, 
an even number. But
one must have $r<2r/a<2r$ which implies that $1<a$ and $a<2$ 
,hence, the  possible periodic words in this case have
$g=2$ as the only feasible case. 
For the case $r$ even, the argument is the same. 
Now take $N=2(r+k)$. Then, $\theta_{-}(N)$
for a lattice with $r$ loops can be understood as being 
equivalent to the $\theta_{-}(N)$ of  
paths for a lattice with $r+k$ loops. 
Working on this lattice and using the previous arguments for the case
$k=0$ we obtain the desired result.

Convergence of the infinite product now follows easily using relations
(3.5) and (3.6).
Call $P_{r,n}$ the partial product in (3.1) 
with $N$ running from $r$ up to $n$. Using relations (3.5)
and (3.6) it's found that 
for $n \geq 2r$
$$
P_{r,n}=\prod_{j=[\frac{n}{2}]+1}^{n}(1-z^{2j})^{\theta_{+}(j,r)}
\eqno(3.7)
$$
In the limit $n \rightarrow \infty$
the infinite product only converges if
$\mid z \mid < 1$ and it converges to $1$.
$\Box$

Relations (3.3-4) above reproduces precisely the results of Ref. 1.

Let's consider now the case where
the  $z_{i}$ are all distinct. For this case, a  formula for
$\theta_{+}$ is not yet available 
exception of the case  $r=2$ given in  
Ref. 1 but still one can prove the following. 
%The proof is similar
%to the one above and for this reason it is not given.

\noindent {\bf Theorem 3.2:}
Given  $r$ and  $N \geq r$, 
the number of equivalence classes of nonperiodic paths of length
$N$ with $+$ and $-$ sign, $ \theta_{+}(m_{1},m_{2},...,m_{r})$ and
$ \theta_{-}(m_{1},m_{2},...,m_{r})$ , respectively,
which traverse loops $i_{1}$, ..., $i_{r}$ a number
of times given by $m_{i_{1}}$, ..., $m_{i_{r}}$, $m_{i}>0$, respectively,
$m_{i_{1}}+...+m_{i_{r}}=N$
the following relations hold:

\noindent {\bf I)} If $m_{i_{1}},...,m_{i_{r}}$ are {\bf a)} all odd, 
{\bf b)} coprime or just of distinct
parity or {\bf c)} all even and $N<2r$, then
$$
\theta_{-}(m_{i_{1}},...,
m_{i_{r}},N)=
\theta_{+}(m_{i_{1}},...,
m_{i_{r}},N)
\eqno(3.8)
$$
\noindent {\bf II)} If $m_{i_{1}},...,m_{i_{r}}$ are all even and $N \geq 2r$,
then
$$
\theta_{-}(m_{i_{1}},...,
m_{i_{r}},N)= \theta_{+}(m_{i_{1}},...,
m_{i_{r}},N)-\theta_{+} \left( \frac{m_{i_{1}}}{2},...,
\frac{m_{i_{r}}}{2},\frac{N}{2} \right)
\eqno(3.9)
$$
Furthemore, for $\mid z_{i} \mid < 1$, $i=1,2,...,r$,
the product (1.6) converges to 1.

\noindent {\bf Proof:} Similar to previous.

\newpage
\section*{IV. Connection with Lie algebras  }

In this section a connection with  
finite dimensional Lie algebras is achieved. 
Firstly, the following is proved.

\noindent {\bf Theorem 4.1.}
The number 
of classes of nonperiodic paths of length $N$
which traverse the lattice  $G_{R}$ counterclockwisely is given by:
$$
\theta(N) = \frac{1}{N} \sum_{g \mid N} \mu (g) R^{\frac{N}{g}}
\eqno(4.1)
$$
\noindent {\bf Proof:} 
The proof below is a joint collaboration with A. L. Maciel.

A statement like the one above can be found in Ref. 4. The goal
here is to prove it using formulas from section 3 and the Appendix.

The number of classes of nonperiodic paths of length $N$ 
which traverse counterclockwisely 
a sublattice of $G_{R}$
with $r$ bonds, $r=1, 2,..., R$,
is given by
$$
\theta_{r} (N) = 
 \sum_{ g \mid N} \frac{ \mu (g)}{g} 
 \sum_{\alpha=1}^{\frac{N }{g}} \frac{1}{\alpha}  
\left( \begin{array}{c} \frac{N }{g} -1 \\ \alpha - 1 \end{array} \right)
r w_{r}(\alpha) 
\eqno(4.2)
$$
where $w_{r}(\alpha)$ is given by $(A.11)$
$$
r w_{r}(\alpha)=\sum_{j=1}^{r}(-1)^{r+j}
\left( \begin{array}{c} r \\ j \end{array} \right)
(j-1)^{\alpha} + (-1)^{\alpha + r}
\eqno(4.3)
$$
Thus, the number of classes of nonperiodic paths of length $N$ 
which traverse counterclockwisely 
the lattice $G_{R}$ is
$$
\theta(N)= \sum_{r=1}^{R}  
\left( \begin{array}{c} R \\ r \end{array} \right)
\theta_{r}(N)
\eqno(4.4)
$$
When $N=1$, $\theta_{r}(1)=0$ if $r>1$, hence, $\theta(1)=R$ which
is in (4.1). Consider now the case when $N \geq 2$.
Upon substitution of (4.2) into (4.4) one gets:
$$
\theta(N)=
\frac{1}{N}\sum_{g \mid N} \mu (g)
 \sum_{\alpha=1}^{\frac{N }{g}}  
\left( \begin{array}{c} \frac{N }{g} \\ \alpha\end{array} \right)
A(R,\alpha)
\eqno(4.5)
$$
where
$$
A(R,\alpha)=
(-1)^{\alpha} \sum_{r=2}^{R}(-1)^{r} 
\left( \begin{array}{c} R \\  r \end{array} \right)
+
\sum_{q=1}^{R} \sum_{p=q}^{R} (-1)^{p+q}
\left( \begin{array}{c} R \\  p \end{array} \right)
\left( \begin{array}{c} p \\  q \end{array} \right)
(q-1)^{\alpha}
\eqno(4.6)
$$
The term with the double summation can be written as
$$
\sum_{q=1}^{R-1} (q-1)^{\alpha}    
\sum_{p=q}^{R} (-1)^{p+q}
\left( \begin{array}{c} R \\  p \end{array} \right)
\left( \begin{array}{c} p \\  q \end{array} \right)
+
(R-1)^{\alpha}
\eqno(4.7)
$$
The summation over $p$
is equal to zero for $q < R$ $^{9}$, so that
$$
A(R,\alpha)=
(-1)^{\alpha}(R-1) + (R-1)^{\alpha}
\eqno(4.8)
$$
and
$$
\theta(N)= \frac{1}{N} \sum_{ g \mid N} \mu (g)
(-R + R^{\frac{N}{g}})
\eqno(4.9)
$$
Using that $\sum_{g | N} \mu(g)=0$  for $N \geq 2$ $^{7}$, the result follows.
$\Box$ 

The right hand side in formula (4.1) is well known in Lie algebra theory
where it is known as the Witt formula $^{7-8}$.
It gives the dimensions of the subspaces 
 $L_{N}$ of a Lie algebra
with a $Z_{>0}$-gradation
$L= \bigoplus_{N=1}^{\infty} L_{N}$ generated by an $R$-dimensional vector
space over ${\bf C}$  $^{7-8}$.
Witt formula satisfies the identity
$$
\prod_{n=1}^{\infty} (1-z^{n})^{\dim L_{n}} = 1- Rz
\eqno(4.10)
$$
called the denominator identity $^{11}$ for the algebra.

Recently, in Refs. 7-8, Witt formula has been generalized.
Let $V= \bigoplus_{i=1}^{\infty} V_{i}$ be a ${\bf Z}_{>0}$-graded vector
space over ${\bf C}$ with $dim V_{i}= d(i) < \infty$, $\forall i \geq 1$,
and let $L= \bigoplus_{N=1}^{\infty} L_{N}$ be the free Lie algebra
generated by $V$ 
with a $Z_{>0}$-gradation induced by that of $V$.
Then, the dimensions of the subspaces $L_{N}$
are given by the generalized Witt formula
$$
dim L_{N}= \sum_{g | N} \frac{\mu(g)}{g} W \left(\frac{N}{g}\right)
\eqno(4.11)
$$
where $W$ is the Witt partition function. $W$ and  the $d(i)$ are related. 
For its definition in terms of
$d(i)$'s, see Refs. 7-8. 
The denominator identity in this case is
$$
\prod_{N=1}^{\infty} (1-z^{N})^{dim L_{N}}= 1- f(z)
\eqno(4.12)
$$
where
$$
f(z)= \sum_{i=1}^{\infty} d(i) z^{i}
\eqno(4.13)
$$
is related to Witt partition function as follows. Define
$$
g(z)=\sum_{n=1}^{\infty} W(n)z^{n}
\eqno(4.14)
$$
Then,
$$
e^{-g(z)}=  1-f(z)
\eqno(4.15)
$$
Comparing (4.11) and (4.2)
it is clear that they have the same general structure.
For each $r$ let's interpret (4.2) as the dimension of some Lie algebra 
$L^{(r)}$. Then the dimensions of the associated vector spaces are given
by the following theorem:

\noindent {\bf Theorem 4.2.} For each $r=1,2,3,...$, 
consider the free Lie algebra
$L^{(r)}= \bigoplus_{N=1}^{\infty} L_{N}^{(r)}$ with 
$dim L_{N}^{(r)}$ given
by (4.11) (or (4.2)) with Witt partition function
$$
W^{(r)} \left( n \right) = \sum_{\alpha=1}^{n} \frac{1}{\alpha}  
\left( \begin{array}{c} n -1 \\ \alpha - 1 \end{array} \right)
r w_{r}(\alpha) 
\eqno(4.16)
$$
and  $rw_{r}$ given by (4.3). 
Then, for $|z|< r^{-1}$, the dimensions $d_{r}(i)= dim V_{i}^{(r)}$
of the vector spaces in $V^{(r)}= \bigoplus_{i=1}^{\infty} V_{i}^{r}$,
the vector space that generates $L^{(r)}$, is given by
$$
d_{r}(i)= \frac{1}{i!} \frac{d^{i}f_{r}}{dz^{i}} (0)
\eqno(4.17)
$$
where
$$
f_{r}(z)= 1 - \prod_{j=1}^{r}(1-jz)^{C_{j}(r)}
\eqno(4.18)
$$
and
$$
C_{j}(r)= (-1)^{j+r}
\left( \begin{array}{c} r \\ j \end{array} \right)
\eqno(4.19)
$$
In particular, for $r=1$, $d(j)=0$ if $j \neq 1$ and $d(1)=1$; for $r=2$, 
$$
d(0)=0, \hspace{5mm} d(j)=j-1, \hspace{5mm} j \geq 1;
\eqno(4.20)
$$
for $r=3$, 
$$
d(0)=d(1)=d(2)=0, \hspace{5mm} d(j)=2^{j-2}(j-2)(j+5), \hspace{5mm} j \geq 3.
\eqno(4.21)
$$
\noindent {\bf Proof:} Take $|z|< r^{-1}$. Using (4.3), relation (4.16)
can be expressed as
$$
W^{(r)}(n) = \sum_{j=1}^{r} (-1)^{j+r}  
\left( \begin{array}{c} r \\ j \end{array} \right)
\frac{j^{n}}{n} 
\eqno(4.22)
$$
and substituting in (4.14), one obtains
$$
g(z)= \sum_{j=1}^{r} (-1)^{j+r+1}
\left( \begin{array}{c} r \\ j \end{array} \right) ln(1-jz)
= ln \prod_{j=1}^{r}(1-jz)^{-C_{j}}
\eqno(4.23)
$$
and from (4.15), follows (4.18).
$f_{r}(z)$ is analytic in $z=0$, so (4.17) holds. For $r=1$, $f_{1}(z)=z$;
for $r=2$,
$$
f_{2}(z)= \frac{z^{2}}{(1-z)^2}
\eqno(4.24)
$$
and for $r=3$,
$$
f_{3}(z)= \frac{2z^{3}- 3z^{4}}{(1-2z)^{3}} 
\eqno(4.25)
$$
Expanding (4.24) and  (4.25) one gets (4.20) and (4.21). $\Box$

From (4.12) the denominator identities are
$$
\prod_{N=1}^{\infty} (1-z^N)^{\theta_{1}(N)}= 1-z
\eqno(4.26)
$$
and
$$
\prod_{N=1}^{\infty} (1-z^N)^{\theta_{2}(N)}= 1 - \frac{z^2}{(1-z)^{2}}
\eqno(4.27)
$$
and
$$
\prod_{N=1}^{\infty} (1-z^N)^{\theta_{3}(N)}= 
1 - \frac{2z^3 - 3z^4}{(1-2z)^{3}}
\eqno(4.28)
$$
for $r=1$, $r=2$ and $r=3$, respectively. (4.26) is trivially true.
One can chek that $\theta_{1}(1)=1$
and $\theta_{1}(N)=0$ if $N\neq 1$

\noindent {\bf Theorem 4.3.} For each $r=2,3,...$, 
consider the free Lie algebra
$L^{(r)}= \bigoplus_{N=1}^{\infty} L_{N}^{(r)}$ with 
$dim L_{N}^{(r)}$ given by (4.11) with Witt partition function 
$W^{r}(n)=2W_{r}(n)$
where 
$$
W_{r}(n)=
 \sum_{l=1}^{n} \frac{1}{l}  
\left( \begin{array}{c} n -1 \\ l - 1 \end{array} \right)
2^{l-1}r w_{r}(l) 
\eqno(4.29)
$$
Then, for $|z|<(2r+1)^{-1}$, the dimensions $d_{r}(i)= dim V_{i}^{(r)}$
of the vector spaces in $V^{(r)}= \bigoplus_{i=1}^{\infty} V_{i}^{r}$,
the vector space that generates $L^{(r)}$, is given by
$$
d_{r}(i)= \frac{1}{i!} \frac{d^{i}f_{r}}{dz^{i}} (0)
\eqno(4.30)
$$
where
$$
f_{r}(z)=1-(1+z)^{(-1)^{r}} \prod_{k=0}^{r-1} 
[1-(2k+1)z]^{-b(k)} 
\eqno(4.31)
$$
with
$$
b(k)= \frac{r(-1)^{r+k}}{(k+1)}
\left( \begin{array}{c} r -1 \\ k \end{array} \right)
\eqno(4.32)
$$
In particular, for $r=2$, 
$$
d(0)=d(1)=0, \hspace{5mm} d(j)= 4(j-1) \hspace{5mm} j \geq 2
\eqno(4.33)
$$
For $r=3$, $d(0)=d(1)=d(2)=0$
and, for $j\geq 3$
$$
d(j)=\frac{6}{8}(-1)^{j-1} + \frac{5}{16} 3^{j-3}[(4j+39)^2 - 43]
-\frac{3^{j-2}}{16}[(4j+13)^2-43]
\eqno(4.34)
$$
\noindent{\bf Proof}: For each $r$ take $|z|< (2r+1)^{-1}$.
Upon substitution of $W(n)$ in (4.14) one gets
$$
g(z)=ln\{ (1+z)^{(-1)^{r+1}} \prod_{k=0}^{r-1} 
[1-(2k+1)z]^{b(k)} \}
\eqno(4.35)
$$
and from (4.15), follows (4.31).
In particular, for $r=2$,
$$
f(z)=1- \frac{(1+z)(1-3z)}{(1-z)^{2}} = \frac{4z^{2}}{(1-z)^{2}}
\eqno(4.36)
$$
and
for $r=3$:
$$
f(z)=1- \frac{(1-z)^{3}(1-5z)}{(1+z)(1-3z)^{3}}
\eqno(4.37)
$$
Expanding (4.36) and (4.37) one gets (4.33) and (4.34). $\Box$

The denominator identity for the algebra in this case is
$$
\prod_{j=1}^{\infty} (1-z^{j})^{\theta (j,r)}= 
(1+z)^{(-1)^{r}} \prod_{k=0}^{r-1} 
[1-(2k+1)z]^{-b(k)} 
\eqno(4.38)
$$

\newpage
\appendix\section*{Appendix A}

Paths of length $N \geq r$
are  described by words of the form
$$
D_{i_{1}}^{e_{i_{1}}}D_{i_{2}}^{e_{i_{2}}}
...D_{i_{l}}^{e_{i_{l}}}
\eqno(A.1)
$$
where  $l=r, r+1, ..., N$, $ i_{k} \neq i_{k+1}, i_{k}=1, ..., r$, and 
$$
\sum_{k=1}^{l} \mid e_{i_{k}} \mid=N
\eqno(A.2)
$$
The number $W_{r}(N)$ of such words
is given by
$$
W_{r}(N)= \sum_{l=r}^{N} 2^{l} p_{l}(N) r w_{r}(l)
\eqno(A.3)
$$
where
$$
p_{l}(N)= \left( \begin{array}{c}
                   N-1 \\ l-1
\end{array}\right)
\eqno(A.4)
$$
is the number of unrestricted partitions of $N$ into $l$ nonzero parts
$\mid e_{i_{k}} \mid$, $k=1, 2, ..., l$. Since
each $e_{i}$ is either positive or negative there are
$2^{l}$ ways of assigning $+$ and $-$ signs to these numbers.
Call $\alpha_{1}, \alpha_{2}, ..., \alpha_{r}$, the loops of $G_{r}$.
Given
$i_{1} \in \{\alpha_{1}, \alpha_{2}, ..., \alpha_{r} \}$,
denote by $w_{r}(l)$ the number of sequences 
$(i_{1}, i_{2}, ...., i_{l})$ with $i_{1}$ fixed and 
$ i_{k} \in \{ \alpha_{1}, \alpha_{2}, ..., \alpha_{r} \}$, 
such that: {\bf a)}
each $\alpha_{k}$ shows up at least once in the sequence;
{\bf b)}
$i_{k} \neq i_{k+1}$;
{\bf c)} $i_{l} \neq i_{1}$.
Since there are $r$ possibilities for $i_{1}$ we multiply
$w_{l}(N)$ by $r$ to get all possible sequences.

In the case $r=1$, $w_{1}(1)=1$.
In order to compute a formula for  $w_{r}(l)$, $r > 1$,
let's drop conditions 
{\bf a)} and {\bf c)} for while.
Then, given $i_{1}$ there are $r-1$ possibilities for $i_{2}$, $r-1$
for $i_{3}$ and so on until $i_{l}$. Thus, $(r-1)^{l-1}$
sequences are formed in this way.
Among these sequences there are
$I_{r}(l)$ sequences with $i_{l}=i_{1}$.
Among the remaining sequences there are $q_{r}(l)$ 
sequences  with $i_{l} \neq i_{1}$  and corresponding to paths which do not traverse
all the $r$ lines of $G_{r}$ and  $w_{r}(l)$
sequences which obey the conditions {\bf a)}, {\bf b)} and {\bf c)}.
It is clear that
$$
w_{r}(l)=(r-1)^{r-1}- q_{r}(l) -I_{r}(l)
\eqno(A.5)
$$
  In the sequel we derive formulas for $I_{r}$ and $q_{r}$.

\noindent{\bf Lemma A.1.}
$$
I_{r}(l)=\frac{(r-1)^{l-1} + (-1)^{l-1}(r-1)}{r}
\eqno(A.6)
$$

    \noindent{\bf Proof :}
It is assumed that condition {\bf b)} is always satisfied,
so  the number $I_{r}(l)$ is  equal to the number  of 
sequences with 
$l-1$ 
elements with
$i_{l-1} \neq i_{1}$.
The latter is  equal to  the total number of sequences with 
$l-1$ elements
given by $(r-1)^{l-2}$ minus the number of those sequences
which terminate with $i_{l-1}=i_{1}$ given by $I_{r}(l-1)$.
Then,
$$
I_{r}(l)=(r-1)^{l-2} - I_{r}(l-1)
\eqno(A.7)
 $$
whose solution is easily found to be relation (A.6). $\Box$

\noindent {\bf Lemma A.2.} Let $r \geq 2$. Then,
$$
q_{r}(l) =\sum_{k=0}^{r-2}
(-1)^{r+k}
\left( \begin{array}{c}
                   r-1 \\ k
\end{array}\right)
\frac{
k^{l}
 +
(-1)^{l}k}
{k+1}
\eqno(A.8)
$$
\noindent{\bf Proof}: Whe $r=2$, $q_{2}(l)=0$. Suppose now $r>2$.
 Let's consider the  sequences 
$(i_{1}, ..., i_{l})$
where only $k$ of its elements, $k < r$, are distinct.
The number of distinct sequences of 
this type which begin with the same $i_{1}$ is given
by $w_{k}(l)$.
Given $i_{1}$, it remains $r-1$ elements in the set 
$\{1,...,r\}$ 
where $k-1$ elements can  be chosen to make the sequence.
Taking $k=2,3,...,r-1$, it follows that
$$
q_{r}(l)=\sum_{k=2}^{r-1}
\left( \begin{array}{c}
                   r-1 \\ k-1
\end{array}\right) w_{k}(l)
\eqno(A.9)
$$

Now, a recurrence relation for $q_{r}(l)$ can be obtained using
(A.5) 
and (A.6). The solution is (A.8) starting with $k=1$. However, the case
$r=2$ can be included allowing  $k=0$. $\Box$ 

\noindent{\bf Lemma A.3.} Given  $r \geq 2 $,
$$
w_{r}(l)=\sum_{k=1}^{r-1} (-1)^{r+k+1}
\left( \begin{array}{c}
                   r-1 \\ k
\end{array}\right) \frac{k^{l}+k(-1)^{l}}{k+1}
\eqno(A.10)
$$

\noindent {\bf Proof:}
Substitute (A.8) into (A.5) and use (A.6). $\Box$

From (A.10) one gets
$$
r w_{r}(l)=\sum_{j=2}^{r}(-1)^{r+j}
\left( \begin{array}{c} r \\ j \end{array} \right)
(j-1)^{l} + (-1)^{l + r}
\eqno(A.11)
$$
valid for $\forall l \geq r$. Notice that (A.11) can be
extended to allow $j=1$ and the case $w_{1}(1)=1$.

\noindent Also, one can extend $w_{r}(l)$ to allow $l<r$ but in this case:

\noindent {\bf Corollary A.1.} Given $r$ and $ l \leq r-1$,
$$
w_{r}(l)=0
\eqno(A.12)
$$
\noindent {\bf Proof:} The case $l=1$ follows trivially from (A.10).
A simple calculation also shows that 
$$
w_{r}(l+1)+w_{r}(l)=\sum_{\beta=0}^{r-1}(-1)^{\beta}(r-1-\beta)^{l}
\left( \begin{array}{c}
                   r-1 \\ \beta
\end{array}\right)
=(r-1)!S(l,r-1)
\eqno(A.13)
$$
where $S(l,r-1)$ is the Stirling number of second kind.
Using induction and that $S(a,b)=0$ for $a<b$ $^{9}$, the result follows.
$\Box$

In the sequel a formula for $\theta_{+}$, relation (3.3) is computed.
Using (A.12), let's rewrite (A.3) as
$$
W_{r}(N)= \sum_{l=1}^{N} 2^{l} p_{l}(N) r w_{r}(l)
\eqno(A.14)
$$
Following the ideas of Ref. 1 denote by
$\overline{W_{r}(l/g,N/g,s/g,T/g)}$ the number of nonperiodic words
plus their circular permutations associated to the numbers
$l/g$, $N/g$, $s/g$, $T/g$. Then,
$$
W_{r}(l,N,s,T)=\sum_{g \mid (l,N,s,T)} 
\overline{W_{r}(\frac{l}{g},\frac{N}{g},\frac{s}{g},
\frac{T}{g})}
\eqno(A.15)
$$
where the summation is over the common divisors  $g$ of $l,N,s$ and $T$.
The term in (A.15) with $g=1$ counts precisely
the number of distinct nonperiodic words
whereas the $g \neq 1$ terms count the periodic ones.

Applying Mobius
inversion formula $^{10}$ it follows that
$$
\overline{W_{r}(l,N,s,T)}=\sum_{g \mid (l,N,s,T)}
\mu(g)W_{r}(\frac{l}{g},\frac{N}{g},\frac{s}{g},
\frac{T}{g})
\eqno(A.16)
$$
where $\mu$ is the Mobius function$^{1,10}$.
A formula for $W_{r}(l,N,s,T)$ can be computed after the
 the following decompositions in (A.3) are made:
$$
2^{l}=\sum_{s=0}^{l}\left( \begin{array}{c}
                   l \\ s
\end{array}\right)
\eqno(A.17)
$$
and
$$
w_{r}(l)=\sum_{T}f_{r}(T,l)
\eqno(A.18)
$$
where $f_{r}(T,l)$ is the number of sequences with the 
same number $T$ of
subsequences. Explicit knowledge of $f_{r}$ will not be necessary.
From $(A.14)$ it follows that
$$
W_{r}(l,N,s,T)=\left( \begin{array}{c}
                   l \\ s
\end{array}\right)
\left( \begin{array}{c}
                   N-1 \\ l-1
\end{array}\right)
r f_{r}(T,l)
\eqno(A.19)
$$
Using
the rule of signs from section II),
$$
\theta_{\pm}(N,r)=\sum_{l,s,T}\frac{\overline{W_{r}(l,N,s,T)}}{l}
\eqno(A.20)
$$
where  the sum runs over all $l,s$ and $T$ 
satisfying the condition that $l+N+s+T $ is odd for $\theta_{+}$ and 
$l+N+s+T $ is even for $\theta_{-}$.

From relations (A.19),
(A.16) and (A.20) it seems to be the case
that an explicit
formula for $f_{r}(T,l)$ is necessary in order to
find $\theta_{\pm}$. The next calculations will show
that it  suffices to know that $f_{r}$ satisfy (A.18). 
Only the particular case of $\theta_{+}$ with  $N$  even will be
considered explicitly. The other
cases follow similar logic of calculation.

Suppose  $N$ is an even number.
The condition that  $N+l+s+T$ be odd requires that

\noindent a) $l,s,T$  odd;\\
b) $l,s$ even and $T$ odd ;\\
c) $l,T$ even  and $s$ odd ;\\
d) $s,T$  even  and $l$ odd.

On the other hand, for  $N+l+s+T$ to be even requires

\noindent a) $l,s,T$ even ;\\
b) $l,s$ odd  and $T$ even ;\\
c) $l,T$ odd  and $s$ even ;\\
d) $s,T$ odd  and  $l$ even.

In the case that  $N$ is odd just swift the words even and
odd in items $a)-d)$ to get the conditions for $N+l+s+T$
to be odd and likewise in items $e)-h)$ to get the conditions
for $N+l+s+T$ to be even.

For $N$  even, 
the number  $\theta_{+}(N,r)$ is computed 
summing over
$l, s, T$ satisfying the conditions $a), b), c)$ and $d)$ above, that is,
$$
\theta_{+}(N,r) = \left[ \sum_{a)} + \sum_{b)} + \sum_{c)} + \sum_{d)} 
                                         \right]
\frac{1}{l} 
  \sum_{g \mid ( l,N,s,T)}
    \mu (g) W_{r}   \left( \frac{l}{g},\frac{N}{g} , 
\frac{s}{g}, \frac{T}{g} 
    \right)
\eqno(A.21)
$$

Let's consider the a)-sum.
First, the sum over the odd
numbers $l$, $s$, $T$ multiples of a given  $g$ is performed,
and then the sum over the $g$'s which divide $N$. 
From (A.14) the possible values
of $l$ odd are
$l=1, 3, ..., N-1$. Among these, the values  
multiples
of a given  $g$ are $l=(2x+1)g$ for $x=0,1,...,\frac{N}{2g}-1$,
hence
the possible values of $s$ are $s=(2y+1)g$ for $y=0,1,...,x$
and $T=(2z+1)g$. The values of $z$ are not important here and in the
calculations below.

Upon substitution of these  into (A.19)  and (A.21) one gets
for the a)-sum: 
$$
 \sum_{odd \hspace{1mm} g \mid N} \frac{ \mu (g)}{g} 
  \sum_{x=0}^{\frac{N }{2g} - 1} \frac{1}{2x+1}  
\left( \begin{array}{c} \frac{N }{g} -1 \\ 2x \end{array} \right)
  \sum_{y=0}^{x}  \left( \begin{array}{c}  2x+1 \\ 2y+1 \end{array} \right) 
 \nonumber \\
 r\sum_{z} f_{r}( 2z+1,2x+1)
\eqno(A.22)
$$
The b)-sum in (A.21) is over the even values of $l, s$ and
odd values of $T$.
The common divisors are among the odd divisors $g$ of $N$ and
$l=g(2x)$, with $x = 1,...,N/2g$, hence,
$s=g(2y)$, with $y =0,1,...,x$ and
$T=g(2z+1)$.
For the b)-sum one gets:
$$
\sum_{odd \hspace{1mm} g \mid N} \frac{ \mu (g)}{g} 
  \sum_{x=1}^{\frac{N}{2g}} \frac{1}{2x}  
\left( \begin{array}{c} \frac{N }{g} -1 \\ 2x-1 \end{array} \right)
  \sum_{y=0}^{x}  \left( \begin{array}{c}  2x \\ 2y \end{array} \right) 
 r\sum_{z} f_{r}( 2z+1,2x)
\eqno(A.23)
$$
The $l, T$ even and  $s$ odd are given by
$l=g(2x)$, with $x=1,2,...,N/2g$,
$s=g(2y+1)$, with $y=0,1,...,x$ and
$T=g(2z)$.
Then, for the c)-sum:
$$
 \sum_{odd \hspace{1mm} g \mid N} \frac{ \mu (g)}{g} 
  \sum_{x=1}^{\frac{N}{2g}} \frac{1}{2x}  
\left( \begin{array}{c} \frac{N}{g} -1 \\ 2x-1 \end{array} \right)
  \sum_{y=0}^{x}  \left( \begin{array}{c}  2x \\ 2y+1 \end{array} \right) 
 r\sum_{z} f_{r}( 2z,2x)
\eqno(A.24)
$$
The last summation in (A.21)  is over the even values
of  $s, T$ and the odd
$l$, namely,
$l=g(2x+1)$, with $x=0,1,...,\frac{N}{2g} -1$;
$s=g(2y)$, with $y= 0,1,...,x$ and
$T=g(2z)$.
For the d) sum:
$$
\sum_{odd \hspace{1mm} g \mid N} \frac{ \mu (g)}{g} 
  \sum_{x=0}^{\frac{N }{2g}-1} \frac{1}{2x+1}  
\left( \begin{array}{c} \frac{N }{g} -1 \\ 2x\end{array} \right)
  \sum_{y=0}^{x}  \left( \begin{array}{c}  2x+1 \\ 2y\end{array} \right) 
 r\sum_{z} f_{r}( 2z,2x+1)
 \eqno(A.25)
$$
Adding (A.22) to (A.25)
and using that
$$
   \sum_{y=0}^{x}  
\left( \begin{array}{c}  2x+1 \\ 2y+1 \end{array} \right) =
  \sum_{y=0}^{x}  \left( \begin{array}{c}  2x+1 \\ 2y \end{array} \right) 
= 2^{2x}
\eqno(A.26)
$$
yields
$$
\sum_{odd \hspace{1mm} g \mid N} \frac{ \mu (g)}{g} 
  \sum_{x=0}^{\frac{N}{2g}-1} \frac{1}{2x+1}  
\left( \begin{array}{c} \frac{N }{g} -1 \\ 2x\end{array} \right)
2^{2x}  \nonumber \\ 
r\sum_{z}
 \left[ f_{r}( 2z+1,2x+1) + f_{r}( 2z,2x+1)   \right]
  \\
  $$
  $$
  =  \sum_{odd \hspace{1mm} g \mid N} \frac{ \mu (g)}{g} 
  \sum_{x=0}^{\frac{N }{2g}-1} \frac{1}{2x+1}  
\left( \begin{array}{c} \frac{N }{2g} -1 \\ 2x\end{array} \right)
2^{2x} \nonumber \\ 
r\sum_{T} f_{r}( T,2x+1)  
\eqno(A.27)
$$
Adding (A.23) to (A.24) yields:
$$
\sum_{odd \hspace{1mm} g \mid N} \frac{ \mu (g)}{g} 
  \sum_{x=1}^{\frac{N }{2g}} \frac{1}{2x}  
\left( \begin{array}{c} \frac{N }{2g} -1 \\ 2x-1 \end{array} \right)
2^{2x-1}r \sum_{T} f_{r}( T,2x)
\eqno(A.28)
$$
At last, adding ( A.28 ) to ( A.27 ): 
$$
\theta_{+}(N,r)  =  \sum_{odd \hspace{1mm} g \mid N }  
\frac{ \mu(g)}{g}   \hspace{11cm} \nonumber \\
\sum_{T} \left[
  \sum_{x=0}^{\frac{N}{2g}-1} \frac{1}{2x+1}  
\left( \begin{array}{c} \frac{N }{g} -1 \\ 2x\end{array} \right)
2^{2x} r f_{r}( T,2x+1) 
+ \right.   \nonumber \\
$$
$$
+\left.  \sum_{x=1}^{\frac{N }{2g}-1} \frac{1}{2x}  
\left( \begin{array}{c} \frac{N}{g} -1 \\ 2x-1 \end{array} \right)
2^{2x-1}r f_{r}( T,2x) \right]
\eqno(A.29)
$$
Now, put the $x$-sums together and
use (A.18) to get:
$$
\theta_{+} (N, r ) = 
 \sum_{odd \hspace{1mm} g \mid N} \frac{ \mu (g)}{g} 
 \sum_{l=1}^{\frac{N }{g}} \frac{1}{l}  
\left( \begin{array}{c} \frac{N }{g} -1 \\ l - 1 \end{array} \right)
2^{l-1}r w_{r}(l) 
\eqno(A.30)
$$
Although obtained for $N$ even, this formula holds for $N$ odd as well.
Using (A.11) and 
performing the summation over $l$ yields result (3.4).
Relations (3.5-6) can be obtained in the
same fashion as above (see Ref. 1, too) or faster as in section 3.

\newpage
%\begin{thebibliography}{99}

\noindent $^{1}$ G. A. T. F. da Costa, J. Math. Phys. {\bf 38}(2), 1014-1034
(1997).

\noindent $^{2}$ G. A. T. F. da Costa and A. L. Maciel,
 Rev. Bras. Ens. Fis. {\bf 25}, 49
(2003).

\noindent $^{3}$ S. Sherman, J. Math. Phys. {\bf 1}, 202 (1960).

\noindent $^{4}$ S. Sherman, Bull. Am. Math. Soc.{\bf 68}, 225 (1962).

\noindent $^{5}$ S. Sherman, J. Math. Phys. {\bf 4}, 1213 (1963).

\noindent $^{6}$ P. N. Burgoyne, J. Math. Phys. {\bf 4}, 1320 (1963).

\noindent $^{7}$ S. J. Kang and M. H. Kim, J. Algebra {\bf 183}, 560 (1996).

\noindent $^{8}$ S. J. Kang, J. Algebra {\bf 204}, 597 (1998).

\noindent $^{9}$ C. Chuan-Cheng and K. Khee-Meng, Principles and Techniques
in Combinatorics (World Scientific, Singapore, 1992).

\noindent $^{10}$ T. M. Apostol, Introduction to Analytic Number Theory
(Springer Verlag, 1986).

%\end{thebibliography}

\end{document}